\documentclass[preprint,12pt,showkeys,showpacs,nofootinbib,floatfix]{revtex4}
\usepackage{amssymb}
\usepackage{multirow}
\usepackage{amsfonts}
\usepackage{amsmath}
\usepackage{graphicx}
\usepackage{subfigure}
\usepackage[dvipdfm,  
            pdfstartview=FitH,
            bookmarksnumbered=true,
            bookmarksopen=true,
            colorlinks,
            pdfborder=001,
            linkcolor=green,
            anchorcolor=green,
            citecolor=red
            ]{hyperref}
\usepackage{graphicx}
\usepackage[toc,page,title,titletoc,header]{appendix}

\begin{document}
\title{Holographic $p$-wave superfluid}
\author{Ya-Bo Wu$^{1}$}
\thanks{E-mail address:ybwu61@163.com}
\author{Jun-Wang Lu$^{1}$}
\author{Wen-Xin Zhang$^{1}$}
\author{Cheng-Yuan Zhang$^{1}$}
\author{Jian-Bo Lu$^{1}$}
\author{Fang Yu$^{1}$}
\affiliation{
$^1$Department of Physics, Liaoning Normal University, Dalian, 116029, People's Republic of  China}
\begin{abstract}
In the probe limit, we numerically construct a holographic $p$-wave superfluid model in the four-dimensional~(4D) and five-dimensional~(5D) anti-de Sitter black holes coupled to a Maxwell-complex vector field.
 We find that, for the condensate with the fixed superfluid velocity, the results are similar to the $s$-wave cases in both 4D and 5D spacetimes. In particular, the Cave of Winds and the phase transition always being of second order take place in the 5D case. Moreover, we find the translating superfluid velocity from second order to first order increases with the mass squared. Furthermore, for the supercurrent with fixed temperature, the results agree with the Ginzburg-Landau prediction near the critical temperature. In addition, this complex vector superfluid model is still a generalization of the SU(2) superfluid model, and it also provides a holographic realization of the $He_3$ superfluid system.
\end{abstract}
\maketitle
\section{Introduction}
Because of the strong-weak correspondence, the gauge-gravity duality opens an important window for us to understand the nature of strongly coupled gauge field theory by studying its weak gravitational duality~\cite{Maldacena1998}. Recently, this holographic duality has been widely applied to condensed systems, especially high temperature superconductors.

The holographic $s$-wave superconductor model was first built in the four-dimensional~(4D) Schwarzschild anti-de Sitter~(AdS) black hole coupled with a Maxwell-complex scalar field~\cite{Hartnoll2008}. Soon after, the holographic $p$-wave and $d$-wave superconductors were respectively realized by  coupling the SU(2) Yang-Mills (YM) gauge field and a charged spin-two field to the AdS black hole~\cite{Gubser2008a,Chen:2010mk}. Thereafter, various superconductor models were constructed, involving backreaction,  different spacetime backgrounds, and the magnetic field, as well as the analytical method~(see, for example, Refs.~\cite{Horowitz11015,Cai:2010zm,Siopsis, QYPan088, Nishioka131,Albash:2008eh,Lu:2013tza,Cai:2014oca,Cai:2014jta}).

Recently, a new holographic $p$-wave superconductor model was constructed by coupling a Maxwell-complex vector (MCV) field into the 4D Schwarzschild AdS black hole~\cite{Cai:2013pda}. Interestingly, the results showed that due to the nonminimal coupling between the vector field and the Maxwell field, the increasing magnetic field can induce superconductor phase transition even without charge density, which is similar to the QCD vacuum instability triggered by the strong magnetic field to develop the $\rho$-meson condensate~\cite{Chernodub:2010qx}. Moreover, the investigation about the response of the external magnetic field on the $p$-wave phase transition also showed that the SU(2) YM model is a special case of the MCV model in five-dimensional~(5D) soliton and 4D Lifshitz black holes~\cite{Cai:2013kaa,Wu:2014dta,Wu:2014lta}. Furthermore, considering the backreaction, the MCV model exhibits the rich phase structures, especially ``retrograde condensation"~\cite{Cai:2013aca,Li:2013rhw,Cai:2014ija}.

As we know, the typical character of superconductivity is the infinity of conductivity, which means that the steady dc current can exist in the superconducting system even without the external electric field.  From the gauge-gravity duality, we know that the asymptotical value of the bulk is dual to the source of the boundary field theory; hence, the current along one direction at the boundary must correspond to a gauge field along the same direction; Motivated by this, a holographic superfluid solution\footnote{Since the bulk U(1) gauge symmetry is dual to the global U(1) symmetry on the boundary field theory, the holographic phase transition precisely models not the superconductor but rather the superfluid. To build the holographic superconductor, we usually assume that the global U(1) symmetry is weakly gauged. Indeed, the dynamical photon in many condensed matter systems is usually ignored due to its very small effect. Therefore,  we can interpret the holographic solution as not only the superfluid, but also the superconductor.} was constructed by performing a deformation of the superconducting black hole, i.e., turning on a spatial component of the gauge field that only depends on the radial coordinate [$A_\mu(r)$]~\cite{Basu:2008st,Herzog:2008he}. In this superfluid model, the leading term of the asymptotical falloff of $A_\mu(r)$ corresponds to the superfluid velocity, while the coefficient of the subleading term corresponds to the supercurrent.  It follows that a second-order superfluid phase transition occurs when the temperature is lowered below a critical value $T_0$ with vanishing superfluid velocity. For each value of temperature below $T_0$, there exists a critical superfluid velocity, beyond which the superfluid phase is broken into the normal phase, and which increases with decreasing temperature.
Meanwhile, there is a special temperature, beyond~(or below) which the superfluid phase transition is second order~(or first order). We call the critical superfluid velocity corresponding to the special temperature as ``translating superfluid velocity''. In addition, in the case of fixed temperature below $T_0$, the supercurrent improves with the increasing superfluid velocity until it reaches its maximum value, which agrees with the prediction of Ginzburg-Landau (GL) theory~\cite{Tinkham1996}.
In addition, the speed of sound versus the temperature was thermodynamically obtained in the case of the vanishing superfluid velocity in Ref.~\cite{Herzog:2008he}. The results showed that near the critical temperature, the behavior is qualitatively similar to that of the superfluid $He_4$. Subsequently, the authors of Ref.~\cite{Amado:2009ts} studied the hydrodynamics of holographic $s$-wave superconductors from the quasinormal mode~(QNM), identified with the poles of the retarded Green function, and found that the speed of the second sound agree with the ones in Ref.~\cite{Herzog:2008he}.

Next, the holographic superfluid models were studied in various aspects~(for example, see Refs.~\cite{Kaminski:2009dh,Amado:2013xya,Amado:2013aea,Arias:2014msa,Arean:2010xd,Arean:2010zw,Zeng:2010fs,Zeng:2012xy, Sonner:2010yx,Kuang:2012xe}). Precisely, from the perspective of the QNM analysis, the holographic model~\cite{Amado:2009ts} was generalized to the ones with a bilinear bulk action and the $U(2)$ symmetry in Refs.~\cite{Kaminski:2009dh,Amado:2013xya,Amado:2013aea,Arias:2014msa}, respectively. In particular, by applying the Landau criterion to the QNM spectrum, the phase diagram was revisited in Ref.~\cite{Amado:2013aea}, where the results displayed a much lower critical temperature compared to the one from the thermodynamical analysis~\cite{Herzog:2008he}, and also suggested that there might exist a spatially modulated phase slightly beyond the critical temperature. On the other hand, many interesting features were also obtained by using the thermodynamical calculations~\cite{Arean:2010xd,Arean:2010zw,Zeng:2010fs,Zeng:2012xy, Sonner:2010yx,Kuang:2012xe}. For example, the holographic superfluid phase transition is always of the second order in the black hole background with a strong enough  backreaction from the matter field~\cite{Sonner:2010yx}, as well as the soliton background at the probe level~\cite{Kuang:2012xe}. In the case of the fixed supercurrent, the superfluid phase transition is always of the first order for any nonzero supercurrent~\cite{Arean:2010xd, Zeng:2010fs,Zeng:2012xy}. In addition, near the critical temperature, the properties of holographic superfluid models in the 4D spacetime are in quite good agreement with the ones of the GL superconducting film. What is more interesting is that  the $s$-wave superfluid model with different mass exhibits different superfluid phase transitions in the 5D AdS black hole~\cite{Arean:2010zw}. In particular, for some intermediate mass scale, when the temperature decreases, the second-order transition occurs before the first-order transition to a new superconducting phase, which is the so-called Cave of Winds. Thus, it is natural to ask whether the Cave of Winds in the $s$-wave superfluid model still exists for some special mass in the $p$-wave superfluid model, as well as whether the MCV model is still a generalization  of the SU(2) model in the superfluid phase transition.

Based on the above motivations, we construct the holographic superfluid model in 4D and 5D AdS black holes coupled to the MCV field in the probe limit.
We find that, for the condensate with fixed superfluid velocity, the results of the $p$-wave superfluid phase transition are similar to the ones of the $s$-wave case in the 4D and 5D AdS black holes~\cite{Arean:2010zw}. In particular, in the 5D case, there is the Cave of Winds for the intermediate mass, and the phase transition is always of the second order for a high enough mass region. Moreover, for the supercurrent with fixed temperature, our results are consistent with not only the ones obtained from the condensate with fixed superfluid velocity, but also the GL prediction near the critical temperature. Furthermore, the MCV model is still a generalization of the  SU(2) YM model in  the holographic superfluid model.

The paper is organized as follows. In Sec.~II, we obtain the equations of motion and the grand potential for the superfluid model in 4D and 5D black holes. Using the shooting method,  we study the vector condensate with fixed superfluid velocity in Sec.~III. The supercurrent versus the superfluid velocity is studied with fixed temperature in Sec.~IV. The final section is devoted to  the conclusions and further discussions.

\section{Equations of motion for holographic superfluid }
In this section, we derive the equations of motion for the matter field of the holographic $p$-wave superfluid phase transition in 4D and 5D black hole spacetimes. To determine the critical temperature in the case of the first-order phase transition, we  obtain the grand potential.

 The $(d+1)$-dimensional Schwarzschild AdS black hole is of the general metric form,
\begin{equation}
ds^2=-r^2f(r)dt^2+\frac{dr^2}{r^2f(r)}+r^2\sum^{d-1}_{i=1}dx_i^2, \qquad f(r)=1-\frac{r_0^d}{r^d},
\end{equation}
where $r_0$ denotes the location of the horizon. The Hawking temperature can easily be written as $T=\frac{d r_0}{4\pi}$. In particular, the 4D~(5D) spacetime corresponds to $d=3$~($4$).

Following Ref.~\cite{Cai:2013aca}, we consider the  matter action  including a Maxwell field and a complex vector field,
\begin{equation}\label{Lvector}
\mathcal{S}_m=\frac{1}{16\pi G_{d+1}}\int dx^{d+1}\sqrt{-g}\left(-\frac{1}{4}F_{\mu\nu}F^{\mu\nu}-\frac{1}{2}\rho_{\mu\nu}^\dag\rho^{\mu\nu}-
m^2\rho^\dag_\mu\rho^\mu+iq \gamma \rho_\mu\rho^\dag_\nu F^{\mu\nu}\right),
\end{equation}
where $F_{\mu\nu}=\nabla_\mu A_\nu-\nabla_\nu A_\mu$ is the strength of the U(1) gauge field $A_\mu$ and $\rho_{\mu\nu}=D_\mu\rho_\nu-D_\nu\rho_\mu$ with $D_\mu=\nabla_\mu-iq A_\mu$, while $m$ ($q$) is the mass (charge) of the vector field $\rho_\mu$.
The interacting  part (characterized by $\gamma$) between the vector field $\rho_\mu$ and the gauge field $A_\mu$ plays an important role in the presence of the external magnetic field~\cite{Cai:2013pda,Cai:2013kaa,Wu:2014dta,Wu:2014lta}. However, we will consider the case without the magnetic field in this paper; hence, it will not contribute to our work.

Moreover, we will proceed with our study on the level of the probe limit, which can be realized by taking $q\rightarrow\infty$ with $q\rho_\mu$ and $qA_\mu$ fixed.  In this probe approximation, the MCV field is taken to be a perturbation for the AdS black hole background; thus, its backreaction to the gravitational background is neglected.

Varying the action (\ref{Lvector}) with respect to the vector field $\rho_\mu$ and the gauge field $A_\mu$ yields the equations of motion
\begin{eqnarray}
 D^\nu\rho_{\nu\mu}-m^2\rho_\mu+iq\gamma\rho^\nu F_{\nu\mu}&=&0,\label{EOMrho}\\
 \nabla^\nu F_{\nu\mu}-iq(\rho^\nu\rho^\dag_{\nu\mu}-\rho^{\nu\dag}\rho_{\nu\mu})
  +iq\gamma\nabla^\nu(\rho_\nu\rho^\dag_\mu -\rho^\dag_\nu\rho_\mu)&=&0.\label{EOMphi}
 \end{eqnarray}

To model the vector phase transition with a superfluid, we turn on the following Ans\"{a}tze:
\begin{equation}\label{rhoA}
\rho_\nu dx^\nu=\rho_x(r) dx,\ \ \ \ \  A_\nu dx^\nu=\phi(r) dt+A_y(r)dy,
\end{equation}
where we have chosen the gauge to fix the phase of $\rho_\mu$ and further taken $\rho_x$ to be real. In the remainder of this paper, we set $q=1$. Comparing with the superconductor model, such as in Refs.~\cite{Cai:2013pda,Wu:2014dta,Wu:2014lta},  what is different is that we have turned on the spatial component of the gauge field $A_y(r)$ to model the superfluid. Substituting the above Ans\"{a}tze (\ref{rhoA}) into Eqs.~(\ref{EOMrho}) and (\ref{EOMphi}) results in the following equations:
\begin{eqnarray}
\rho_x ''+\left(\frac{f'}{f}+\frac{d-1}{r}\right)\rho_x' -\frac{ \rho_x}{r^2 f}\left(m^2+\frac{A_y^2}{r^2}-\frac{\phi ^2}{r^2 f}\right) &=&0, \label{Eqpsi}\\
\phi ''+\frac{d-1}{r}\phi '-\frac{2 \rho_x ^2 }{ r^4 f}\phi&=&0,\label{Eqphi}\\
A_y''+\left(\frac{ f'}{f}+\frac{d-1}{r}\right)A_y'-\frac{2 \rho_x ^2}{ r^4 f}A_y&=&0,\label{EqAx}
 \end{eqnarray}
where the prime denotes the derivative with respect to $r$.
From Eq.~(\ref{Eqpsi}), we get the effective mass of the vector field,
 \begin{equation}\label{meff2}
 m^2_{eff}=m^2-\frac{\phi ^2}{r^2 f} +\frac{A_y^2}{r^2},
 \end{equation}
 which implies that the increasing $m^2$ and $A_y^2(r)$ will hinder the superfluid phase transition, while the increasing $\phi(r)$ will enhance the transition. Generally speaking, it is very difficult to solve the above equations analytically, but the numerical method is still feasible~(for example, the shooting method~\cite{Herzog:2008he,Basu:2008st,Sonner:2010yx,Kuang:2012xe,Arean:2010xd,Zeng:2010fs,Zeng:2012xy,Arean:2010zw}). To solve Eqs.~(\ref{Eqpsi}), (\ref{Eqphi}), and (\ref{EqAx}), we impose some boundary conditions. At the horizon, $\rho_x(r_0)$ and  $A_y(r_0)$ should be regular, while the gauge potential $A_t(r_0)$ must vanish in order for $g^{\mu\nu}A_\mu A_\nu$ to remain finite. At the infinity boundary $r\rightarrow\infty$, the general expansions of the fields are of the forms
\begin{equation}\label{asyrpa}
\rho_x(r)=\frac{\rho_{x-}}{r^{\Delta_-}}+\frac{\rho_{x+}}{r^{\Delta_+}}+\cdots,\quad
\phi(r)=\mu-\frac{\rho}{r^{d-2}}+\cdots,\quad
A_y(r)=S_y-\frac{J_y}{r^{d-2}}+\cdots
\end{equation}
with the characteristic exponent $\Delta_\pm=\frac{1}{2}\left(d-2\pm\sqrt{(d-2)^2+4m^2}\right)$. According to the gauge-gravity duality, we usually interpret the coefficients of the most dominant term ($\rho_{x-}$) and the subleading term ($\rho_{x+}$) as the source  and the vacuum expectation value of the boundary operator $O_x$, respectively, while $\mu$, $\rho$, $S_y$, and $J_y$ are understood as the chemical potential, the charge density, the superfluid velocity, and the supercurrent, respectively. In particular, in the case of the Breitenlohner-Freedman~(BF) bound, i.e., $m^2_{BF}=-\frac{(d-2)^2}{4}$, a logarithmic term will emerge in the general falloff of $\rho_x(r)$. In order to avoid the instability, we take the coefficient of the logarithmic term as the source and the one of the other term as the vacuum expectation value.

There exists an important scaling symmetry for the asymptotical solutions~(\ref{asyrpa}), which reads
 \begin{equation}
 (r,r_0,T,\mu,S_y)\rightarrow \lambda (r,r_0,T,\mu,S_y),\qquad \rho_{x+}\rightarrow \lambda^{\Delta_++1} \rho_{x+},\qquad (\rho, J_y)\rightarrow \lambda^{d-1} (\rho, J_y),
 \end{equation}
 where $\lambda$ is always a positive real constant. This symmetry is also corret for the logarithmic case of the vector field $\rho_{x}$, which we can use to fix the chemical potential $\mu$ and the superfluid velocity $S_y$, and thus work in the grand canonical ensemble. From Refs.~\cite{Herzog:2008he,Basu:2008st,Arean:2010zw}, in the grand canonical ensemble, the order of phase transition will change from second order to first order when one increases the superfluid velocity beyond the translating value. In this case, to determine the critical temperature and which phase is more thermodynamically favored, we should calculate the grand potential $\Omega$ of the bound state. According to the gauge-gravity duality, $\Omega$ is identified by the product of the Hawking temperature and the on-shell action with the Euclidean signature. From the action (\ref{Lvector}), we obtain the on-shell action $\mathcal{S}_{os}$ as
\begin{eqnarray}
\mathcal{S}_{os}&=&\int dx^{d-1}dtdr\sqrt{-g}\left(-\frac{1}{2}\nabla_\mu (A_{\nu}F^{\mu\nu})-\nabla_\mu(\rho^\dag_{\nu}\rho^{\mu\nu})+\frac{1}{2}A_\nu\nabla_\mu F^{\mu\nu}\right)\nonumber\\
&=&\frac{V_{d-1}}{T}\left(-\sqrt{-\gamma}n_r\rho^\dag_{\nu}\rho^{r\nu}|_{r\rightarrow\infty}-\frac{1}{2}
\sqrt{-\gamma}n_r A_\nu F^{r\nu}|_{r\rightarrow\infty}+\frac{1}{2}\int_{r_0}^\infty dr \sqrt{-g}A_\nu\nabla_\mu F^{\mu\nu}\right)\nonumber\\
&=& \frac{V_{d-1}}{T}\left(\frac{d-2}{2}(\mu\rho-S_y J_y)+\int_{r_0}^\infty dr r^{d-5}\rho_{x+}^2\left(A_y^2-\frac{\phi^2}{f}\right)\right),
\end{eqnarray}
where we have taken advantage of the expansions of the gauge field $A_t$ and $A_y$, and also considered the integration $\int dt dx^{d-1}=\frac{V_{d-1}}{T}$, as well as neglected the prefactor $\frac{1}{16\pi G_{d+1}}$ for simplicity. It should be stressed that since we work in the probe limit, we do not need to introduce the Gibbons-Hawking boundary term for the well-defined Dirichlet variational problem. In addition, under the source-free boundary condition for $\rho_x$, we find that there is no divergent term in the on-shell action; therefore, we have not added the counterterm. For the sake of the numerical calculation, we usually work in the new coordinate $z=\frac{r_0}{r}$; hence, the grand potential in the superfluid phase $\Omega_S$ reads
\begin{eqnarray}\label{FreeE}
\frac{\Omega_{S}}{V_{d-1}}&=&- \frac{T\mathcal{S}_{os}}{V_{d-1}} =\frac{d-2}{2}(-\mu\rho+S_y J_y)+\int_{\epsilon}^1 dz z^{3-d}\rho_{x+}^2\left(\frac{\phi^2}{1-z^d}-A_y^2\right),
 \end{eqnarray}
 where the lower bound $z\rightarrow\epsilon$ corresponds to the boundary $r\rightarrow\infty$.

 In the normal phase, i.e., $\rho_x(r)=0$, the solutions for Eqs.~(\ref{Eqphi}) and (\ref{EqAx}) are given by
 \begin{equation}
 \phi(r)=\mu(1-z^{d-2}),\qquad\qquad A_y(r)=S_y,
 \end{equation}
 where we have considered the finite form of $\phi(z)|_{z=1}$ and $A_y(z)|_{z=1}$ as in Refs.~\cite{Basu:2008st,Zeng:2010fs}; therefore, the grand potential in the normal phase $\Omega_N$ is of the form
 \begin{equation}
 \frac{\Omega_{N}}{V_{d-1}}=-\frac{d-2}{2}\mu^2.
 \end{equation}
\section{Results of condensates with fixed superfluid velocity}
In this section, we calculate the condensate for some different values of mass beyond the BF bound and superfluid velocity in the 4D and 5D spacetimes, respectively. To get the condensate, we also calculate the corresponding grand potential. In particular, we consider the following cases of mass in detail, which are listed in Table~\ref{Delta}.
\begin{table}[htbp]
\centering
\caption{\label{Delta}$\Delta_-$, $\Delta_+$, and the critical temperature without the superfluid velocity for the different values of mass in the 4D and 5D AdS black holes.}
\begin{tabular}{|c|c|c|c|c|c|c|c|c|}
\hline
$d$ & \multicolumn{4}{|c|}{$d=3$} & \multicolumn{4}{|c|}{$d=4$}\\
\hline
$m^2$&$-\frac{1}{4}$&$0$ & $\frac{3}{4}$&$2$ &$-1$&  $0$& $\frac{5}{4}$ & $3$\\
\hline
$\Delta_-$&$\frac{1}{2}$&0&$-\frac{1}{2}$&$-1$&$1$&$0$&$-\frac{1}{2}$&$-1$ \\ \hline
$\Delta_+$&$\frac{1}{2}$&1&$\frac{3}{2}$&$2$&$1$&$2$&$\frac{5}{2}$&$3$ \\ \hline
$\frac{T_0}{\mu}$&$0.1237$&$0.0654$&$0.0437$&$0.0326$&$0.1785$&$0.0796$&$0.0614$&$0.0498$\\
\hline
\end{tabular}
\end{table}

\subsection{Results of condensates in the 4D case}
\begin{figure}
\begin{minipage}[!htb]{0.45\linewidth}
\centering
\includegraphics[width=3.0in]{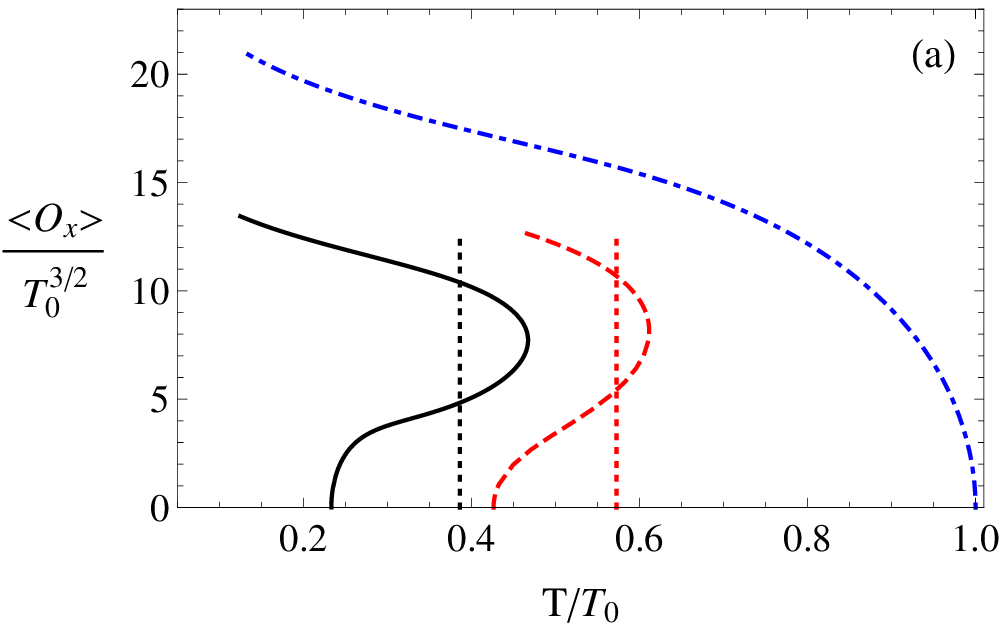}
\end{minipage}
\begin{minipage}[!htb]{0.45\linewidth}
\centering
\includegraphics[width=3.0in]{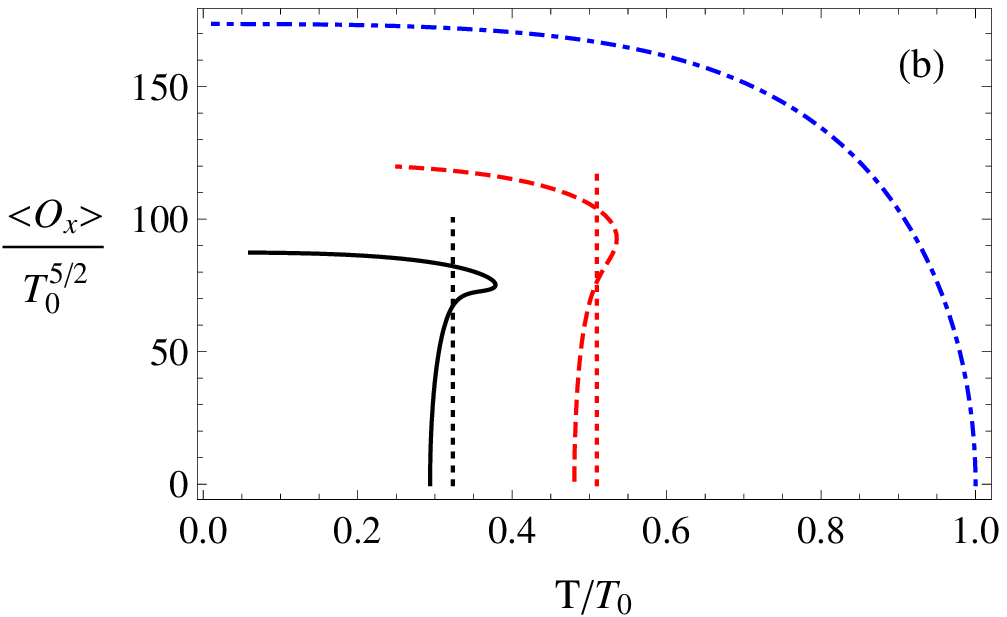}
\end{minipage}
\begin{minipage}[!htb]{0.45\linewidth}
\centering
 \includegraphics[width=3.0 in]{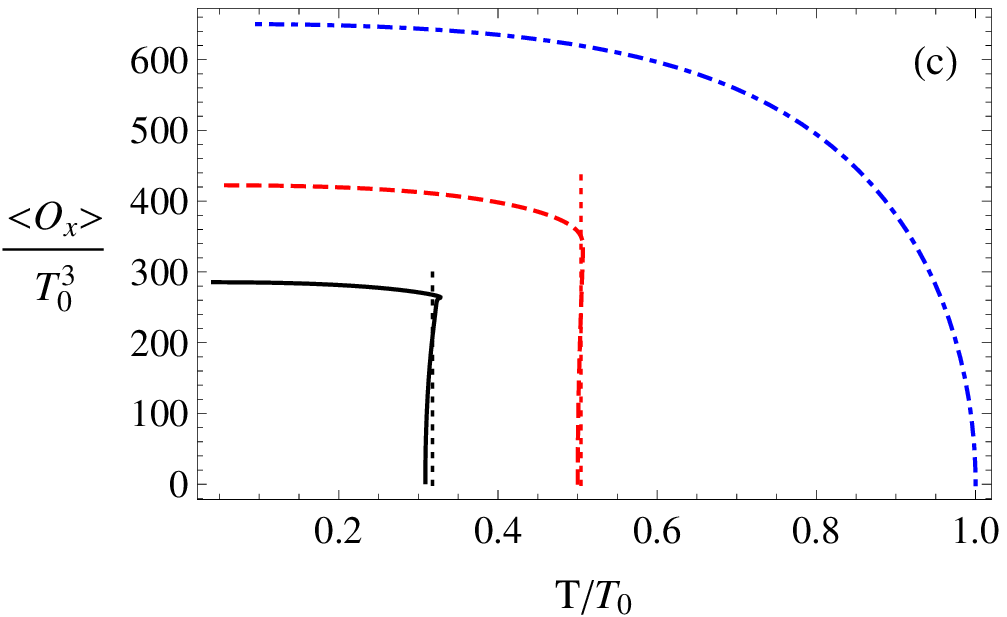}\\
\end{minipage}
\begin{minipage}[!htb]{0.45\linewidth}
\centering
 \includegraphics[width=3.0 in]{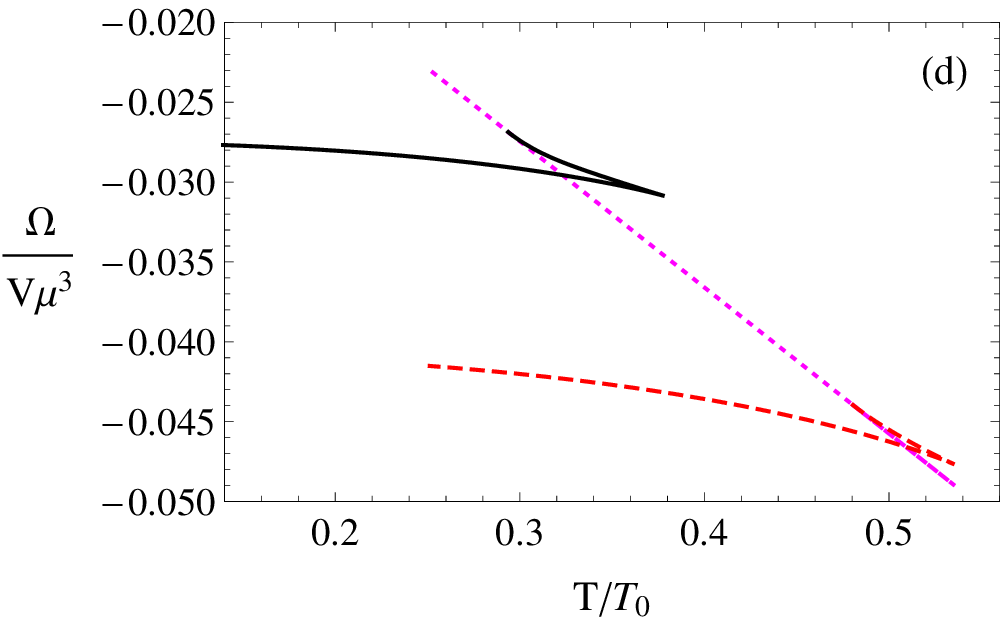}\\
\end{minipage}
\caption{The condensate versus the reduced temperature with $m^2=-\frac{1}{4}$~(a),~$\frac{3}{4}$~(b), and 2~(c), where all curves respectively correspond to $\frac{S_y}{\mu}=0,~0.5$, and $0.65$ from right to left. Panel (d) illustrates the grand potential in the case of $m^2=\frac{3}{4}$ corresponding to $\frac{S_y}{\mu}=0.5$ (red dashed line), $0.65$ (black solid line), and the normal phase (magenta dotted line), respectively. }
\label{d4ConGrand}
\end{figure}

Using the shooting method and the boundary conditions mentioned above, we numerically solve Eqs.~(\ref{Eqpsi}), (\ref{Eqphi}), and (\ref{EqAx}) and then  plot the condensate and the grand potential of the 4D background in Fig.~\ref{d4ConGrand}, where we have not plotted the case of $m^2=0$ because it is similar in behavior to that of $m^2=\frac{3}{4}$.
The rightmost curves in Figs.~\ref{d4ConGrand}(a), \ref{d4ConGrand}(b), and \ref{d4ConGrand}(c) denote the superconducting phases without superfluid velocity (i.e., $S_y=0$), from which we find that, when the temperature gradually decreases, there is always a critical value below which the second-order phase transition occurs. This is just the conductor-superconductor phase transition~\cite{Cai:2013pda,Wu:2014dta,Cai:2013aca}. The critical temperature $T_0$ with $S_y=0$ for different values of mass is listed in Table~\ref{Delta}; it follows that the increasing mass squared $m^2$ makes the phase transition more difficult, which is clear from the effective mass of the vector field $\rho_x(r)$, i.e., Eq.~(\ref{meff2}), where the increasing mass improves the effective mass and thus hinders it from decreasing below the BF bound. In particular, in the case of $m^2=\frac{3}{4}$, the results are in accordance with the ones in Refs.~\cite{Cai:2013pda,Wu:2014dta} by rescaling the unit. For the small superfluid velocity, with decreasing temperature, the phase transition is still of the second order until the superfluid velocity improves to a special value, beyond which the phase transition changes from the second order to the first order.

We plot the condensate with  some high enough superfluid velocity for different values of mass in Figs.~\ref{d4ConGrand}(a), \ref{d4ConGrand}(b), and \ref{d4ConGrand}(c), i.e., the curves with $\frac{S_y}{u}=0.5$ and $0.65$. Obviously, the condensate versus the temperature becomes double valued at some temperature, which indicates the first-order transition.  To determine the critical temperature of the first-order phase transition, we should calculate the grand potential $\Omega$ for the superconducting phase and the normal phase. Here we representatively plot the grand potential in the case of $m^2=\frac{3}{4}$ corresponding to $\frac{S_y}{u}=0.5$ (red dashed line) and $0.65$ (black solid line) as well as the normal phase (magenta dotted line) in Fig.~\ref{d4ConGrand}(d), from which we find that there exist clear swallowtails, which are a remarkable signal of the first-order transition. Moreover, by comparing the grand potential in the superfluid phase with the one in the normal phase, we mark the locations of the critical temperature  with a vertical dotted line in the same color as the condensate curve. It follows that the larger the superfluid velocity, the smaller the critical temperature, which is obvious from the effective mass~(\ref{meff2}), where the increasing $A_y(r)$ improves the effective mass and thus hinders the phase transition.

\begin{figure}
\includegraphics[width=2.9in]{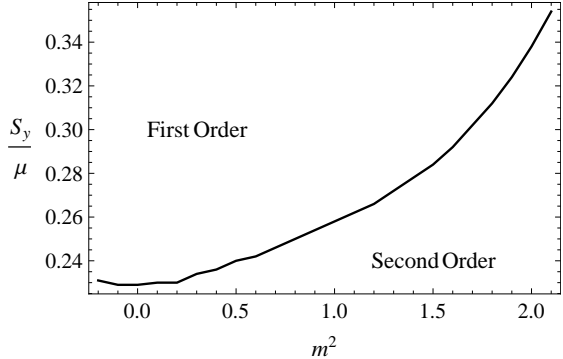}
\caption{The translating superfluid velocity $\frac{S_y}{\mu}$ from the second to the first order as a function of the mass squared of the vector field $\rho_\mu$. }
\label{d4TranP}
\end{figure}

To clearly see the effects of the mass squared on the phase transition, we plot the dependence of the translating superfluid velocity $\frac{S_y}{\mu}$ on the mass from the second order to the first order in Fig.~\ref{d4TranP}.
Now, we must determine which side of the bound curve denotes the first-order phase transition. From Figs.~\ref{d4ConGrand}(a), \ref{d4ConGrand}(b), and \ref{d4ConGrand}(c), we know that if one further improves the superfluid velocity, the superfluid phase transition will change from the second order to the first order; therefore, the upper left region denotes the first order while the other part represents the second-order transition. It follows that the translating value of $\frac{S_y}{\mu}$ almost improves monotonously with the increasing mass squared $m^2$; i.e., the larger mass squared hinders the emergence of the translating point. It can be understood from the limit of the large mass, where the contribution from $A_y^2(r)$ to the effective mass can be ignored, that the superfluid model reduces to the superconductor model, which is always of the second order in the probe limit. In our cases, however, we can always realize the first-order phase transition in the 4D case, which agrees with the results of Refs.~\cite{Herzog:2008he,Basu:2008st,Arean:2010zw}.

As discussed in the works of superconductors, for example, in Refs.~\cite{Cai:2013kaa,Wu:2014dta,Wu:2014lta}, the MCV model is a generalization of the SU(2) YM model in the standard AdS black hole, even in Lifshitz and Gauss-Bonnet gravity. Comparing our Eqs.~(\ref{Eqpsi}), (\ref{Eqphi}), and (\ref{EqAx}) with the ones~[i.e., Eqs.~(2.8), (2.9), and (2.10) in Ref.~\cite{Zeng:2010fs}] obtained from the SU(2) YM model, it is evident that the equations of motion are identical to each other by rescaling the vector field $\rho_x(r)$. Therefore, we conclude that the MCV model is still a generalization of the SU(2) YM model in the holographic superfluid model.

\subsection{Results of the condensate in the 5D case}
 From calculations in the case of different mass squared, we find that the results of our MCV superfluid model are similar to the case of the $s$-wave superfluid model in the 5D AdS black hole~\cite{Arean:2010zw}. In particular, for small mass beyond the BF bound, the features of the phase transition in the 5D spacetime are similar to the case of the 4D spacetime. For the intermediate mass scale, the Cave of Winds appears, and for a high enough mass, the phase transition is always of the second order.

\begin{figure}
\begin{minipage}[!htb]{0.45\linewidth}
\centering
\includegraphics[width=3.0in]{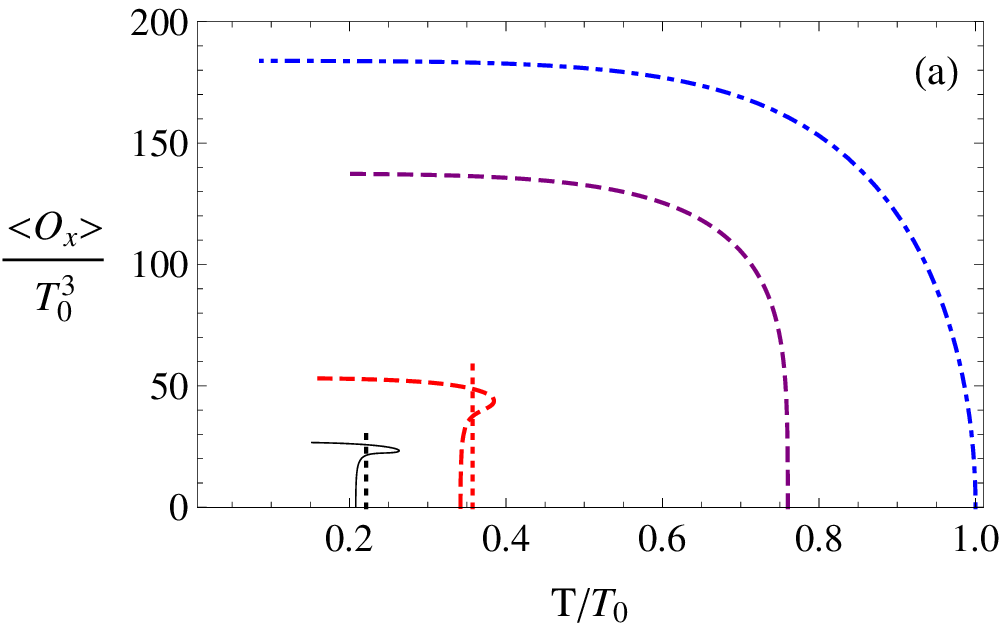}
\end{minipage}
\begin{minipage}[!htb]{0.45\linewidth}
\centering
 \includegraphics[width=3.0 in]{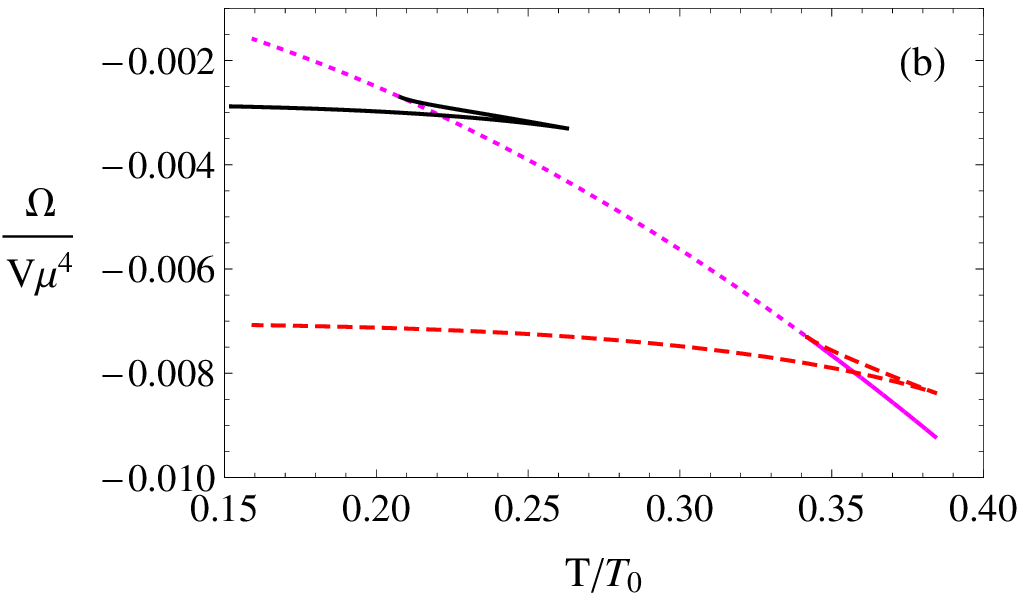}\\
\end{minipage}
\caption{The condensate (a) and grand potential (b) versus the reduced temperature with $m^2=0$ in the 5D black hole. The curves in (a) correspond to $\frac{S_y}{\mu}=0$, $0.42$, $0.75$, and $0.85$ from right to left,  while the curves in (b) to $\frac{S_y}{\mu}=0.75$ (red dashed), $0.85$ (black solid), and the normal phase (magenta dotted), respectively. }
\label{dim5m0ConGran}
\end{figure}

We plot the condensate and the corresponding grand potential with $m^2=0$, as an example of the small mass scale, in Fig.~\ref{dim5m0ConGran}.
In the case of vanishing or small superfluid velocity (i.e., $\frac{S_y}{\mu}=0$ and $0.42$), the second-order phase transition is triggered as the temperature is lowered below a critical value. With the increasing superfluid velocity (i.e., $\frac{S_y}{\mu}=0.75$ and $0.85$), the phase transition switches from the second order to the first order, which is the result we can clearly see from the grand potential with the typical swallowtails in Fig.~\ref{dim5m0ConGran}(b). The critical temperature for the first-order phase transition is also marked by the vertical dotted line from the calculation of the grand potential. Moreover, the critical temperature decreases with the increasing superfluid velocity, which agrees with GL theory. Furthermore, we calculate the value of the translating superfluid velocity $\frac{S_y}{\mu}$ from the second order to the first order in the small mass region, and we find that it almost increases monotonously with $m^2$ ($-1\leq m^2 \leq\frac{1}{2}$), which is similar to Fig.~\ref{d4TranP} and thus indicates that the larger $m^2$ hinders the translation.

\begin{figure}
\begin{minipage}[!htb]{0.45\linewidth}
\centering
\includegraphics[width=3.0in]{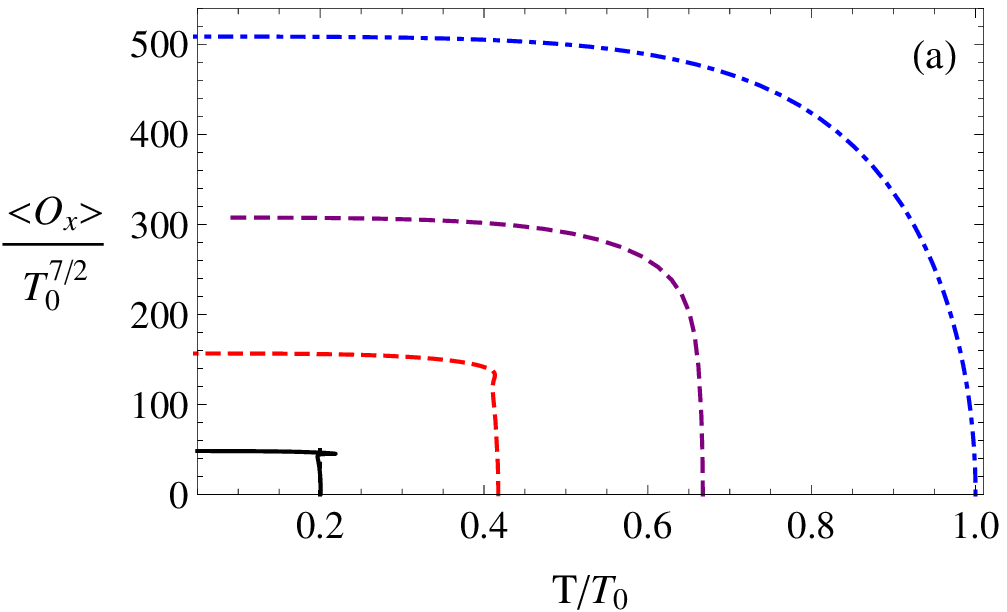}
\end{minipage}
\begin{minipage}[!htb]{0.45\linewidth}
\centering
 \includegraphics[width=3.0 in]{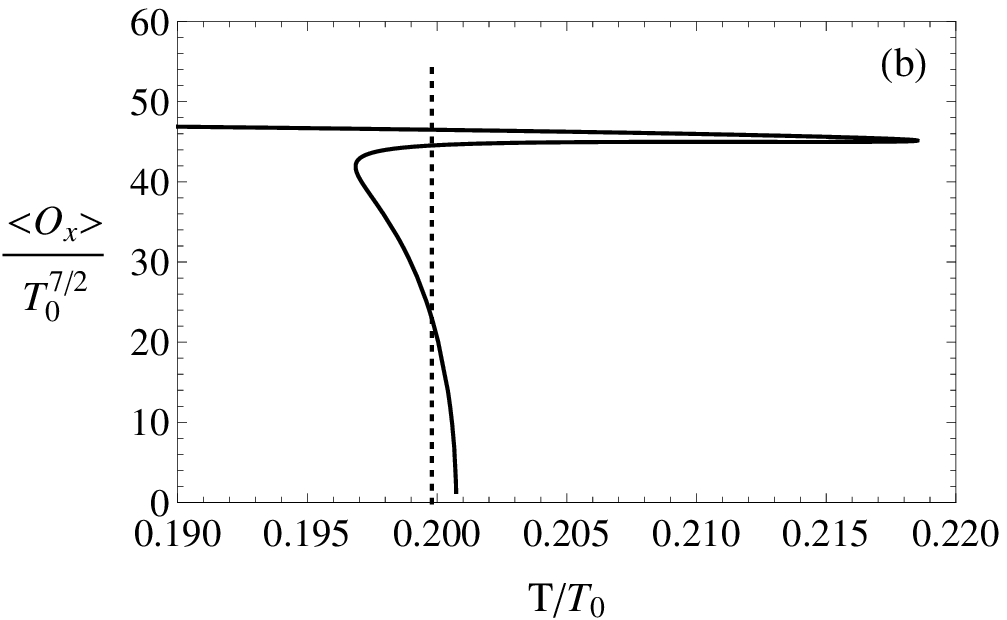}\\
\end{minipage}
\begin{minipage}[!htb]{0.45\linewidth}
\centering
 \includegraphics[width=3.0 in]{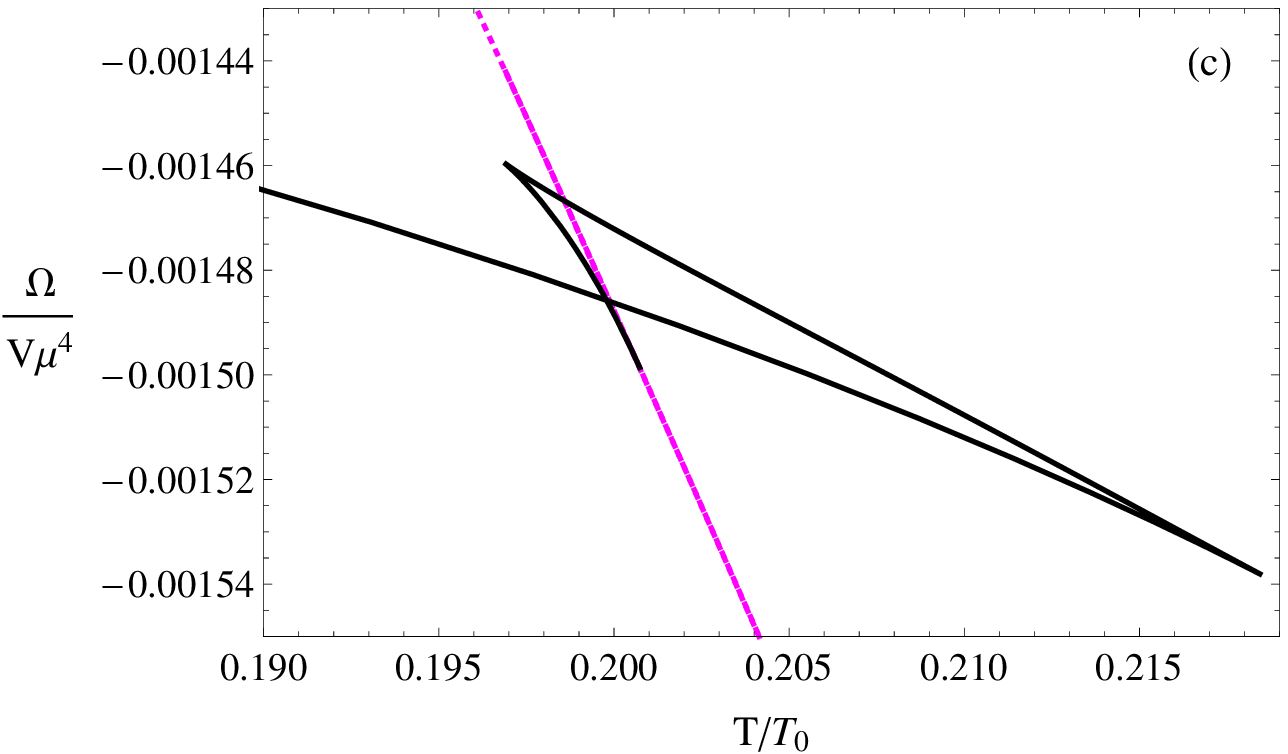}\\
\end{minipage}
\caption{The condensates (a), (b) and the grand potential (c) as a function of the reduced temperature with $m^2=\frac{5}{4}$ in the 5D black hole. The curves in (a) correspond to $\frac{S_y}{\mu}=0$, $0.5$, $0.7$, and $0.86$  from right to left, and the curve in (b) to $\frac{S_y}{\mu}=0.86$, while ones in (c) to $\frac{S_y}{\mu}=0.86$ (black solid) and the normal phase (magenta dotted), respectively.}
\label{d5m5f4ConGran}
\end{figure}

What is more interesting is the case of the intermediate mass.  As an example, we  plot the condensate and the grand potential in the case of $m^2=\frac{5}{4}$ in Fig.~\ref{d5m5f4ConGran}.
It follows that, for small enough superfluid velocity (i.e., $\frac{S_y}{\mu}=0$ and $0.5$), the phase transition is still of second order in the process of decreasing temperature. However, when the superfluid velocity increases to a special value, the first-order phase transition will follow from the second-order phase transition. From the grand potential in Fig.~\ref{d5m5f4ConGran}(c), we notice the following: Since the grand potential stretches smoothly from the normal phase and is lower than that of the normal phase, the system first suffers a second-order phase transition and is dominated by the superfluid phase. Then, the grand potential crosses the curve of the normal phase two times. However, even the grand potential is lower than that of the normal phase; it does not have the lowest potential, so it is not thermodynamically favorable.  Only  when the curve crosses the curve of the second-order phase transition is the system thermodynamically stable. We show this transition point by the black vertical dotted line in the condensate; i.e., the lowest-right and the highest-left parts of the condensate in Fig.~\ref{d5m5f4ConGran}(b) are thermodynamically stable.  This interesting phase diagram is the so-called Cave of Winds, which is similar to the case of the Maxwell-complex scalar model~\cite{Arean:2010zw} and will appear in the GL theory when some higher-order terms that are more than quartic in the action change the sign.

\begin{figure}
\begin{minipage}[!htb]{0.45\linewidth}
\centering
\includegraphics[width=3.0in]{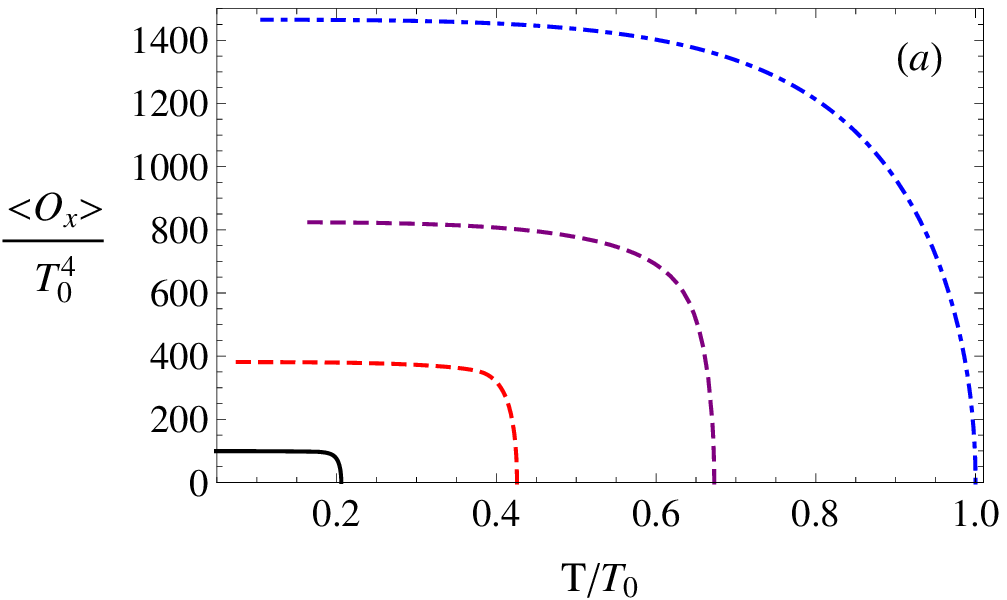}
\end{minipage}
\begin{minipage}[!htb]{0.45\linewidth}
\centering
 \includegraphics[width=3.0 in]{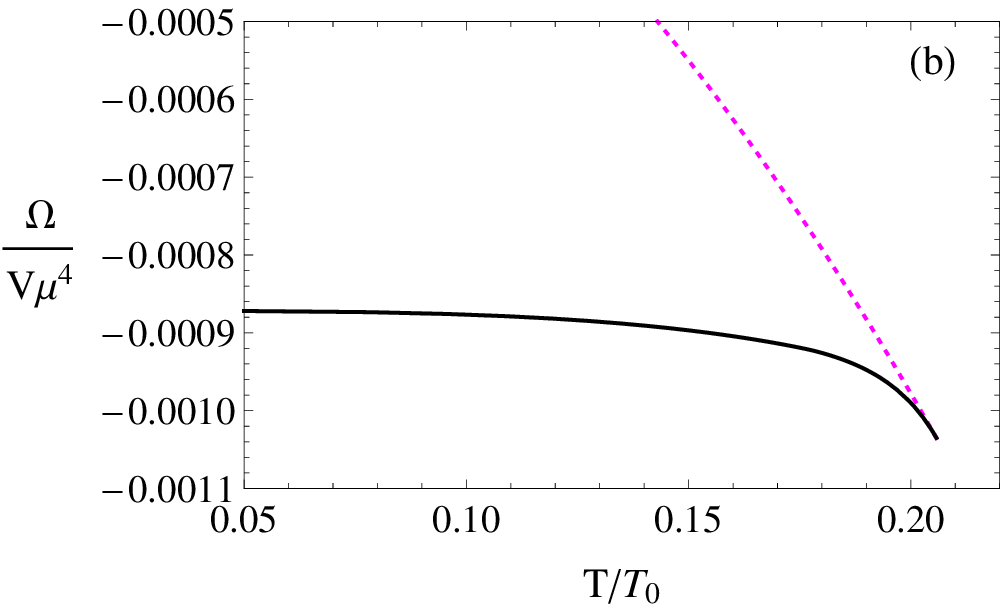}\\
\end{minipage}
\caption{The condensate (a) and the grand potential (b) versus the reduced temperature  with $m^2=3$ in the 5D black hole. The curves in  (a) correspond, respectively, to $\frac{S_y}{\mu}=0$, $0.5$, $0.7$, and $0.86$ from right to left, while the ones in (b) correspond to $\frac{S_y}{\mu}=0.86$ (black solid) and the normal phase (magenta dotted). }
\label{d5m3ConGrand}
\end{figure}

The condensate for a high enough mass squared  is the most interesting, and it is plotted with $m^2=3$ in Fig.~\ref{d5m3ConGrand}.
From the figure, we find that even in the rather high superfluid velocity $\frac{S_y}{\mu}=0.86$, when the temperature decreases, the system still suffers not a first-order phase transition but rather the second-order one, which is given by the corresponding grand potential in Fig.~\ref{d5m3ConGrand}(b).

As we know from Refs.~\cite{Hartnoll2008,Gubser2008a,Lu:2013tza,Cai:2013pda,Wu:2014dta,Wu:2014lta}, in the absence of the spatial component of the gauge field $A_y(r)$, i.e., the superconductor model, the system always suffers the second-order phase transition in the probe limit. However, with the presence of $A_y(r)$ in the superfluid model, we find that the phase transition will switch from the second order to the first order in the case of a sufficiently high superfluid velocity. Therefore, $A_y(r)$ is crucial to the order of the phase transition. From Eq.~(\ref{meff2}), we see that $A_y(r)$ appears in the effective mass, and in the case of a high enough mass squared $m^2$, the effect of $\frac{A^2_y(r)}{r^2}$ becomes relatively so weak that it can be ignored; therefore, we always obtain the second-order phase transition in the case of a high mass squared in the 5D black hole. In the case of $m^2=2$ in the 4D spacetime, we still observe the first-order phase transition when the superfluid velocity increases sufficiently, but it is evident that we need the larger superfluid velocity to realize the translating point from the second order to the first order in the case of a higher mass squared than the case of a lower mass squared.

Since the SU(2) YM model is a generalization of the MCV model, the interesting condensate versus the temperature in the 5D MCV model will also appear in the SU(2) YM model. In particular, in the case of the fixed supercurrent, we can obtain results similar to the ones in Ref.~\cite{Zeng:2010fs}.
\section{Results of supercurrents with fixed temperature}
\begin{figure}
\begin{minipage}[!htb]{0.45\linewidth}
\centering
\includegraphics[width=2.8in]{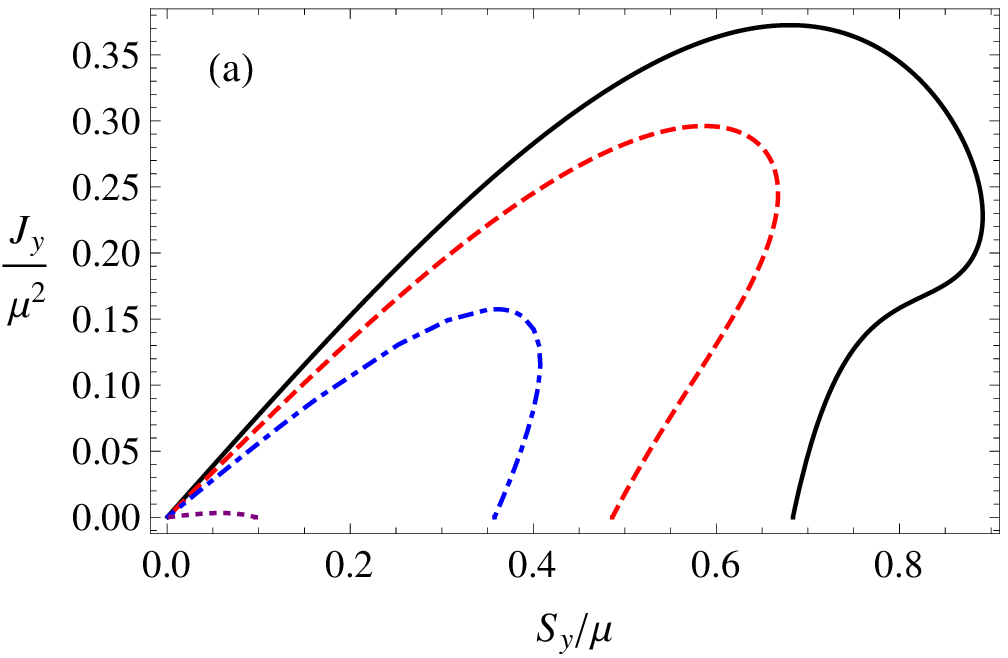}
\end{minipage}
\begin{minipage}[!htb]{0.45\linewidth}
\centering
 \includegraphics[width=2.8 in]{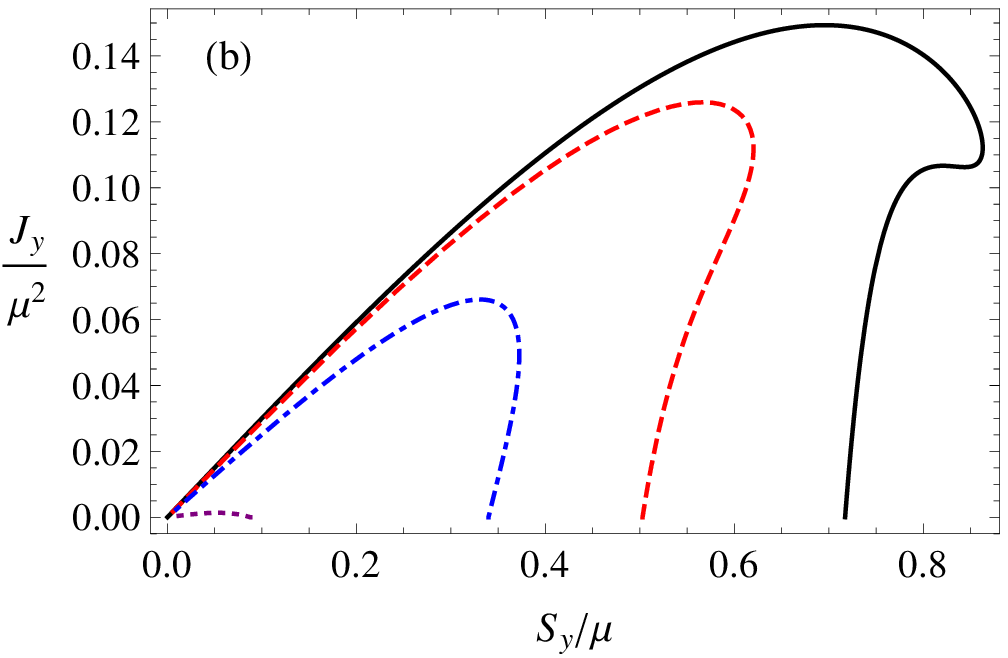}\\
\end{minipage}
\begin{minipage}[!htb]{0.45\linewidth}
\centering
\includegraphics[width=2.8in]{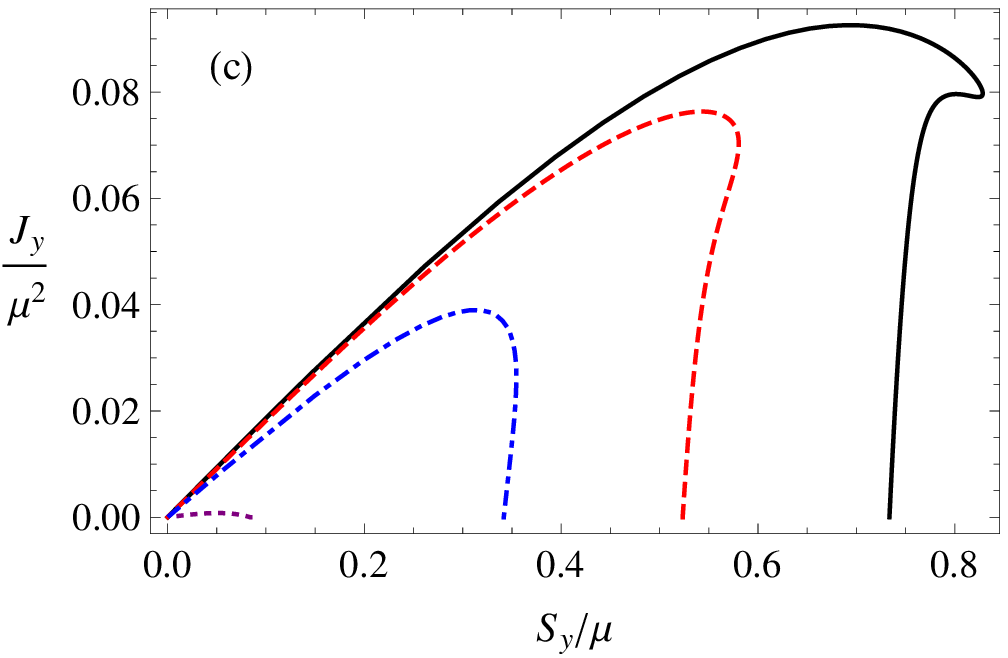}
\end{minipage}
\begin{minipage}[!htb]{0.45\linewidth}
\centering
 \includegraphics[width=2.8 in]{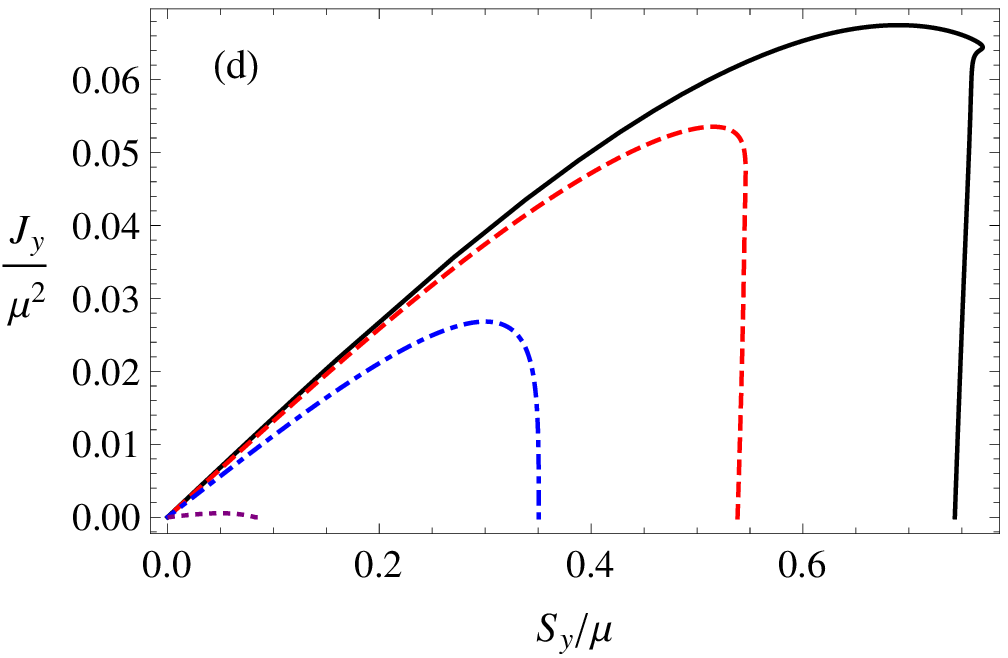}\\
\end{minipage}
\caption{The supercurrent versus superfluid velocity for $m^2=-\frac{1}{4}$~(a),~0~(b),~$\frac{3}{4}$~(c), and 2~(d) in the 4D case.  The curves in all panels  correspond to $\frac{T}{T_0}=0.2,~0.45,~0.7$, and $0.98$ from right to left, respectively. }
\label{d4CurrVFlu}
\end{figure}

As we know, near the critical temperature, the GL theory may give an exact description for the superconductor, such as the thin superconducting film. To check the reasonability of our holographic model, it is helpful to compare our results with the prediction of the GL theory. Since our 4D gravitational black hole geometry can directly correspond to the 2D superconducting film, we first plot  the relation between the supercurrent and the superfluid velocity, with fixed temperature in the 4D case, in Fig.~\ref{d4CurrVFlu}, where  the mass squared is chosen with the calculations of Sec.~III.

From Fig.~\ref{d4CurrVFlu}, we see that
for all cases of the mass squared, near the critical temperature, the curves (such as $\frac{T}{T_0}=0.98$) approximate a parabola opening downward, and the maximum value of the supercurrent $\frac{J_y}{\mu^2}$, which is denoted by $\frac{J_{yMax}}{\mu^2}$, decreases with the increasing reduced temperature. All the curves have two intersecting points with the abscissa axis. The first one is at the location  where the superfluid velocity $\frac{S_y}{\mu}$ vanishes. It is reasonable from Eq.~(\ref{asyrpa}) that $S_y$ stands for the source of the supercurrent $J_{y}$. The other point is denoted by $\frac{S_{yMax}}{\mu}$, beyond which the superconducting phase breaks into the normal phase, and it can be understood from the effective mass that the spatial component $A_y(r)$ hinders the superfluid phase transition.
With the increase of $\frac{S_y}{\mu}$ from zero to $\frac{S_{yMax}}{\mu}$, $\frac{J_{y}}{\mu^2}$ improves  up to its maximum value $\frac{J_{yMax}}{\mu^2}$ and then decreases smoothly to zero at  $\frac{S_{yMax}}{\mu}$, which implies the second-order phase transition in accordance with the results in Sec.~III, and also agrees with GL thoery~\cite{Tinkham1996,Sonner:2010yx,Arean:2010xd,Zeng:2010fs,Zeng:2012xy}. Moreover, for a fixed supercurrent less than $\frac{J_{yMax}}{\mu^2}$, there exist two velocities. As shown in Ref.~\cite{Arean:2010xd}, the state with the smaller superfluid velocity has a lower grand potential, and thus is thermodynamically favorable.

In the case of  the temperature evidently deviating from the critical temperature, the linear dependence of the supercurrent on the superfluid velocity becomes more obvious until its maximum value $\frac{J_{yMax}}{\mu^2}$, which matches the one in the thin superconducting films~\cite{Tinkham1996}; and the larger the mass squared, the smaller the maximum value $\frac{J_{yMax}}{\mu^2}$, which is in agreement with the fact that the increasing mass hinders the phase transition. Moreover, when the superfluid velocity improves beyond the larger intersecting point with the abscissa axis,\footnote{It should be noted that when the temperature is lowered sufficiently below the critical temperature, the larger intersecting point is no longer the maximum value of the superfluid velocity.} the supercurrent versus the superfluid velocity becomes double valued.  For this case, if we further improve beyond the maximum value of the superfluid velocity $\frac{S_y}{\mu}$,
 the supercurrent will jump from a nonzero value to zero, which must result in the latent heat, and thus assigns the first-order phase transition~\cite{Sonner:2010yx,Arean:2010xd,Zeng:2010fs,Zeng:2012xy}.  In addition, the larger the mass squared $m^2$, the more difficult it is for the curve of the supercurrent to become double valued, which again suggests that the larger $m^2$  hinders the emergence of the translating superfluid velocity from the second-order to the first-order phase transition.  Meanwhile, according to the previous condensate in Fig.~\ref{d4ConGrand}, for fixed reduced temperature and the superfluid velocity, if  there exist two values, the state with the larger condensate has the lower grand potential; hence, we conclude that the state with the larger supercurrent is more thermodynamically stable. However, this strange structure is qualitatively different from GL theory~\cite{Tinkham1996}.

\begin{figure}
\begin{minipage}[!htb]{0.45\linewidth}
\centering
\includegraphics[width=2.8in]{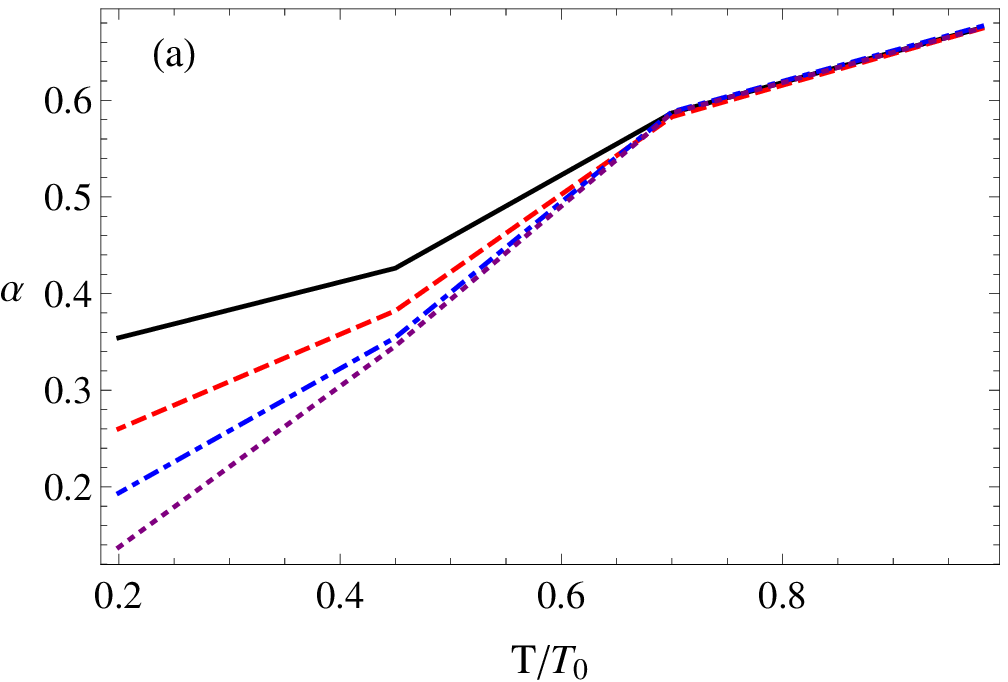}
\end{minipage}
\begin{minipage}[!htb]{0.45\linewidth}
\centering
\includegraphics[width=2.8in]{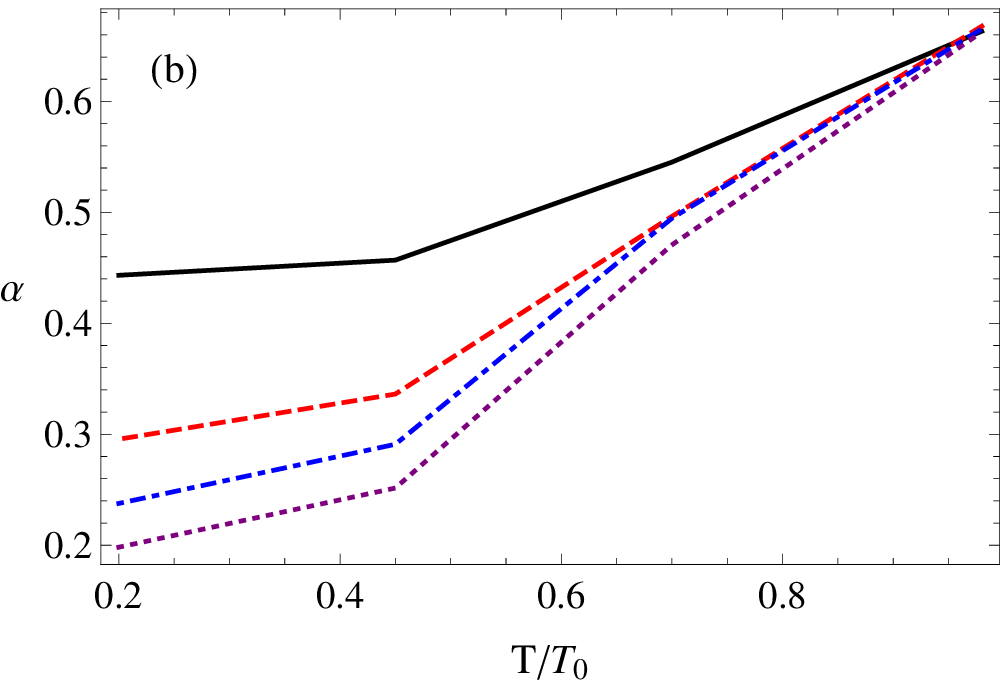}
\end{minipage}
\caption{The ratio $\alpha:=\left(\langle O_x \rangle_c/{\langle O_x \rangle_\infty}\right)^2$ versus the reduced temperature. The curves from top to bottom correspond to the mass squared $m^2=-\frac{1}{4},~0,~\frac{3}{4}$, and $2$ in the 4D spacetime (a), and to $m^2=-1,~0,~\frac{5}{4}$, and $3$ in the 5D case (b), respectively.}
\label{d45ratio}
\end{figure}

As GL theory predicted, at any fixed temperature, the stable value of the condensate decreases monotonically as the superfluid velocity increases. If we denote $\langle O_x \rangle_\infty$ and  $\langle O_x \rangle_c$ as the values of the condensate corresponding to the vanishing superfluid velocity and the one with $\frac{J_{yMax}}{\mu^2}$, respectively, GL theory predicts the following relation
\begin{equation}
\alpha:=\left(\frac{\langle O_x \rangle_c}{\langle O_x \rangle_\infty}\right)^2=\frac{2}{3}.
\end{equation}
We calculate the ratio $\alpha$ for the different values of mass and reduced temperature, and plot the results in Fig.~\ref{d45ratio}(a), from which we find that, near the critical temperature, the ratio is consistent with GL theory for all cases of mass squared. However, the ratio $\alpha$ evidently deviates more from the predicted value when the temperature decreases gradually from the critical temperature and  the mass squared increases from the BF bound $m^2_{BF}=-\frac{1}{4}$.

\begin{figure}
\begin{minipage}[!htb]{0.45\linewidth}
\centering
\includegraphics[width=2.8in]{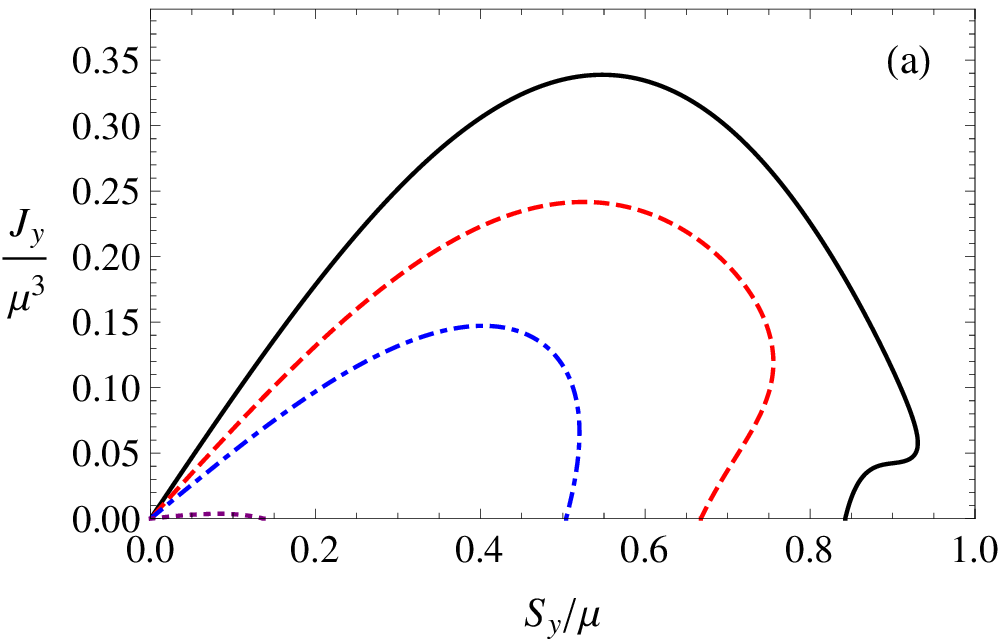}
\end{minipage}
\begin{minipage}[!htb]{0.45\linewidth}
\centering
\includegraphics[width=2.8in]{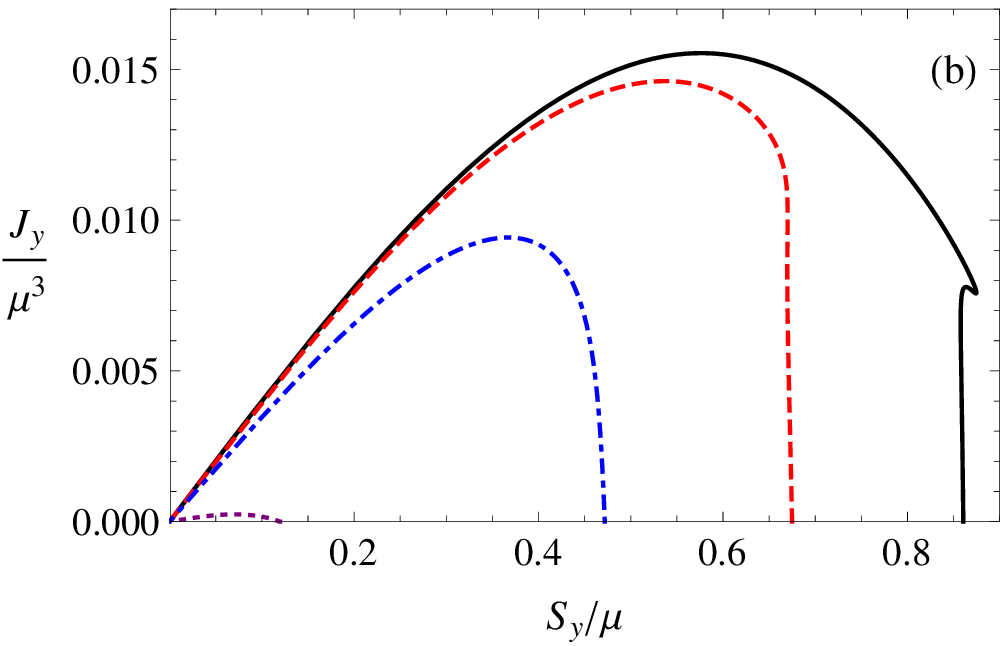}
\end{minipage}
\begin{minipage}[!htb]{0.45\linewidth}
\centering
 \includegraphics[width=2.8 in]{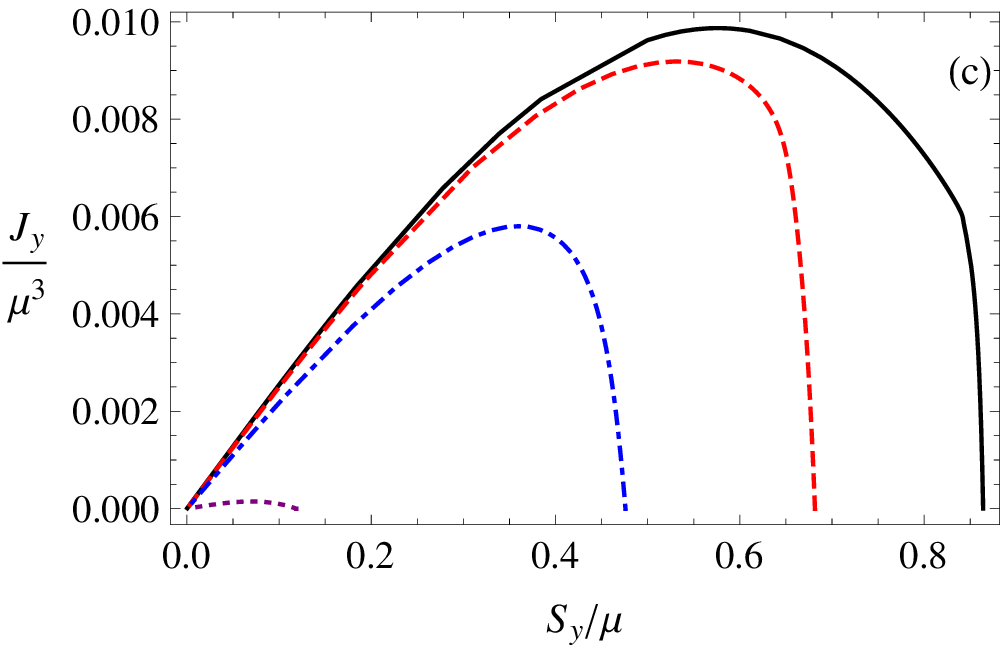}\\
\end{minipage}
\begin{minipage}[!htb]{0.45\linewidth}
\centering
 \includegraphics[width=2.8 in]{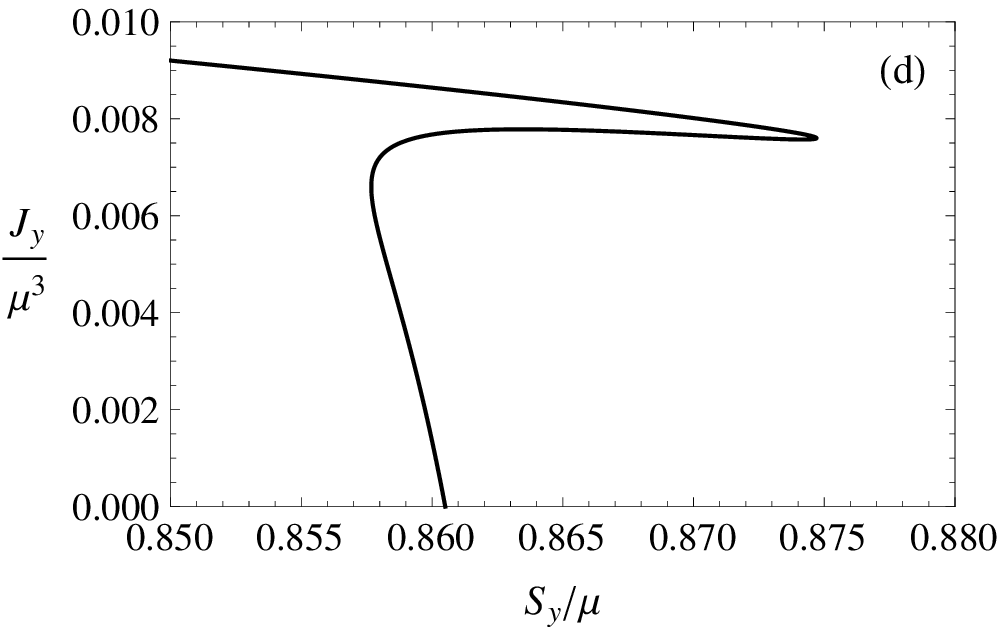}\\
\end{minipage}
\caption{The supercurrent versus superfluid velocity for $m^2=$-1~(a),~$\frac{5}{4}$~(b),~3~(c), and $\frac{5}{4}$~(d) in the 5D spacetime.  The curves in (a), (b), and (c), respectively, correspond to $\frac{T}{T_0}=0.2,~0.45,~0.7$, and $0.98$ from right to left, while the curve in (d) corresponds to $\frac{T}{T_0}=0.2$. Note that the case of $\frac{T}{T_c}=0.98$ corresponds to the lowest purple dotted line.}
\label{d5CurrVFlu}
\end{figure}

Apart from the 4D case, we also plot the supercurrent versus the superfluid velocity for different values of mass squared in the 5D black hole in Fig.~\ref{d5CurrVFlu},
where we have not displayed the case of $m^2=0$, because of its similar behaviors to the case of the BF bound.  From the figure, we find that, for  a small enough mass in Fig.~\ref{d5CurrVFlu}(a), the relation between the supercurrent and the superfluid velocity is similar to the one in the 4D case; i.e., near the critical temperature ($\frac{T}{T_0}=0.98$), the phase transition is of the second order. When the superfluid velocity increases and the reduced temperature decreases to a certain value, the phase transition goes from the second order to the first order ($\frac{T}{T_0}=0.2,~0.45$, and $0.7$), which is in accordance with the  corresponding condensate versus the reduced temperature in Sec.~III. For the intermediate mass, from Fig.~\ref{d5CurrVFlu}(b), we see clearly that, near the critical temperature ($\frac{T}{T_0}=0.98$), the supercurrent versus the superfluid velocity obeys the parabola opening downward, which implies the second-order phase transition. When the reduced temperature goes over a value, the system first suffers a second-order transition and then has a first-order phase transition ($\frac{T}{T_0}=0.2$), which is clearly illustrated by Fig.~\ref{d5CurrVFlu}(d) and agrees with the quantities of the condensate as a function of the reduced temperature in the previous section, i.e., the Cave of Winds in Ref.~\cite{Arean:2010zw}. For the high enough mass in Fig.~\ref{d5CurrVFlu}(c), the system always behaves as the second-order phase transition at the reduced temperatures considered in this paper, which is in agreement with Ref.~\cite{Arean:2010zw}. In addition, we also find that the larger the mass squared $m^2$, the lower the maximum value of the supercurrent $\frac{J_{yMax}}{\mu^3}$, which means that the larger mass hinders the phase transition.

Finally, we calculate the ratio $\alpha$ by taking different values of the mass squared in the 5D black hole, and we plot the results in Fig.~\ref{d45ratio}(b). It follows that near the critical temperature, the value of the ratio $\alpha$ agrees well with the GL predicted value $\frac{2}{3}$. However, when the temperature deviates from the critical temperature, the ratio $\alpha$ decreases. The larger the mass squared, the more obviously the ratio $\alpha$ deviates from $\frac{2}{3}$, similar as those in the 4D case.

\section{Conclusion and discussion}
In this paper, we have numerically studied holographic superfluid models in both 4D and 5D AdS black holes in the probe limit, and we have observed the effects of the mass squared of the vector field  on the superfluid phase transition. The main results are as follows.

For the condensate with fixed superfluid velocity, as the temperature is gradually lowered, the systems always suffer a second-order phase transition when the superfluid velocity vanishes or is very small, whether for all values of mass (4D) or a small mass (5D). For a fixed mass, the critical temperature decreases with  increasing superfluid velocity, which means that the spatial component of the gauge field to modeling the superfluid hinders the phase transition. When the superfluid velocity increases to the translating value, the phase transition changes from the second order to the first order, and the larger the mass squared, the larger the translating superfluid velocity $\frac{S_y}{\mu}$  becomes, which means that the higher mass hinders the emergence of the translating point. In the case of the intermediate mass in the 5D spacetime with a high enough superfluid velocity, the system will suffer a second-order phase transition followed by a first-order one, which is similar to the Cave of Winds of the $s$-wave superfluid model in the 5D AdS black hole~\cite{Arean:2010zw}. In the high mass case of the 5D spacetime, the contribution from the gauge field $A_y(r)$ becomes so weak that it can be ignored, and it just reduces to the superconductor model; therefore, the system is always in the second-order phase transition, independent of the superfluid velocity.

For the supercurrent with fixed reduced temperature, we have found that the results are consistent with the ones obtained from the condensate with fixed superfluid velocity, especially  the Cave of Winds for the intermediate mass and the phase transition being always of the second order for a high enough mass in the 5D black hole. What is more, our results agree with the predictions of the GL theory near the critical temperature in both 4D and 5D spacetimes, in particular, the ratio $\alpha=\left(\langle O_x \rangle_c/{\langle O_x \rangle_\infty}\right)^2$. However, for the fixed reduced temperature, $\alpha$ deviates much more obviously from the GL predicted value  with the increasing mass squared $m^2$. In addition, at the low reduced temperature, the linear dependence of the supercurrent on the superfluid velocity becomes obvious, which matches with the Bardeen-Cooper-Schrieffer~(BCS) superconducting films.

Furthermore, by comparing the equations of motion of the MCV field with the ones of the SU(2) YM field, we have found that the MCV model is still a generalization of the SU(2) YM model in the case of the holographic superfluid model; hence, we conclude that the rich results of the MCV model in the 5D case, such as the Cave of Winds, will also appear in the SU(2) YM model.

In summary, the results of the $p$-wave superfluid phase transition in the 4D and 5D AdS black holes are similar to the ones of the $s$-wave superfluid phase transition~\cite{Arean:2010zw}. Meanwhile, it has been shown that the $He_3$ superfluid system is a $p$-wave model~\cite{Vollhardt1990}; thus, our $p$-wave model might provide a holographic realization for the $He_3$ superfluid in the condensed system in some sense.

It is worth stressing that in this paper, our study of the MCV model has been constrained in the probe limit and compared with the $s$-wave superfluid phase transition, but  the mass boundaries could not be determined in the 5D spacetime. As we know, once the strong backreaction is taken into account, the $s$-wave superfluid phase transition is always of the second order for all superfluid velocities~\cite{Sonner:2010yx}, and the superconductor phase transition of the MCV model exhibits the rich phase structures~\cite{Cai:2013aca,Li:2013rhw,Cai:2014ija}. Thus, as a complementarity, it is valuable to study the backreaction in the superfluid model by coupling the MCV field to the AdS black hole in order to explore whether the rich phase structures, especially in the 5D spacetime still hold in the strong backreaction. In addition, in our work, both the critical temperature and the translating superfluid velocity are determined by comparing the grand potential in the normal phase with the one in the superfluid phase.  As was pointed out in Refs.~\cite{Herzog:2008he,Amado:2013aea}, the superfluid by itself is a metastable state, so the grand potential analysis might not be totally clear. What is more obviously is that by applying the Landau criterion to the QNM spectrum in the case of a linear perturbation with a small momentum, the authors of Ref.~\cite{Amado:2013aea} obtained a much lower critical temperature $T^\ast$ than $\tilde{T}$, where $\tilde{T}$ is the critical temperature from the free-energy analysis. Meanwhile, the fact that the state in the range $ T^\ast<T<\tilde{T}$ becomes instable at finite momentum indicates that there is a spatially modulated phase, namely, the striped phase. It follows that the absence of the related QNM analysis might be a shortcoming of our present work. Therefore, in the following work, we will try our best to study these potentially unstable QNM in our MCV model, which will not only make up for the shortage of this paper, but will also  shed light on the understanding of this $p$-wave superfluid model.

\acknowledgments We would like to thank R.~G. Cai for his directive help. J.~W. Lu is  deeply grateful to L.~Li for his helpful discussions and comments. This work is supported by the National Natural Science Foundation of China (Grants No.~11175077 and No.~11205078), the Ph.D Programs of the Ministry of China (Grant No.~20122136110002), the Project of Key Discipline of Theoretical Physics of the Department of Education in the Liaoning Province (Grants No.~905035 and No. 905061), and the Open Project Program of the State Key Laboratory of Theoretical Physics, Institute of Theoretical Physics, Chinese Academy of Sciences, China (No.~Y4KF101CJ1).

\end{document}